\documentstyle[preprint,tighten,aps,prl]{revtex}
\begin{document}
\draft
\preprint{UTPT-94-04}
\title{A Non-singular Theory of Gravity}
\author{ Neil J. Cornish and John W. Moffat}
\address{Department of Physics, University of Toronto \\ Toronto,
Ontario M5S 1A7, Canada}
\maketitle
\begin{abstract}
We present a geometrical gravitational theory which
reduces to Einstein's theory for weak gravitational potentials
and which has a singularity-free analog of the Schwarzschild
metric.
\end{abstract}
\narrowtext

Despite its geometrical elegance, Einstein's General Theory\cite{Al1} is
plagued by singular solutions which destroy global predictability. It has long
been hoped that a new, non-singular theory of gravity might be found, from
which Einstein's theory emerges as a low-energy limit. We shall present the
static spherically symmetric solution to a close relative of Einstein gravity,
which suggests that we may have found such a theory.

The essential new physics comes from endowing the gravitational field
with additional degrees of freedom. This additional freedom
gives the field the ability to smooth out the energy density of
a point source over a small but finite region, thus removing any
singular behaviour.

The theory we shall study was first put forward as a theory of gravity
in 1979\cite{Moff79}, after it was realized that the field equations
of Einstein-Straus Unified Field Theory\cite{Al} did not describe a unification
of electromagnetism and gravity, but rather a generalized theory of gravity.
Unlike General Relativity, the metric tensor in this theory is not assumed to
be symmetric. This means that both the manifold and the tangent space
transform under $GL(4,R)$, while in General Relativity the tangent space
is restricted to the Lorentz sub-group, $SO(3,1)$, of $GL(4,R)$.
The resulting theory was called Nonsymmetric Gravitational Theory (NGT).

The phenomenology of NGT which has been described in the literature
to-date\cite{banff} turns out to be a singular subset of the full theory.
Both General Relativity and the previously studied sector of NGT emerge as
singular limits of the complete theory.

We shall present an exact, non-singular analog of the Schwarzschild solution
for NGT.
The spacetime this solution describes has no event horizons and no curvature
singularities. The solution set is parameterized by two constants of
integration, $m$ and $s$. We can identify $m$ as the gravitational mass by
taking the Newtonian limit of the solution. The second constant, $s$, is
a dimensionless coupling constant which we shall call the skewness parameter,
since it controls the size of the skew-metric contributions. We shall see that
in the weak field limit, the skew sector of the theory behaves as $sm^2/r^2$
for large $r$. The non-linear coupling to mass emphasizes the purely
geometrical nature of this new sector.

For strong fields we shall see that the limit $s\rightarrow 0$ is non-analytic.
This means that there are no event horizons or singularities for any $s>0$,
while there are both horizons and curvature singularities when $s=0$. That is
what we mean when we say that General Relativity (and the old solutions to NGT)
are singular limits of the full theory.

The most general solution to the field equations for NGT introduces a third
constant of integration, $l^2$. This constant is known as the NGT charge, and
has the dimensions of a length squared. Since this charge arises from a
conserved current, it has been postulated that it is proportional to conserved
particle number. All previous work on NGT has been based on this charge being
non-zero. To simplify our current presentation, we shall take the NGT charge to
be zero. This results in no essential loss of generality since the full theory
is non-singular for arbitrary $l^2$, so long as $s>0$. Moreover, this choice
emphasizes the fact that no new matter couplings are needed to obtain
non-singular solutions.

The NGT Lagrangian without sources is given by\cite{banff}
\begin{equation}\label{NGT}
{\cal L}=\sqrt{-g}g^{\mu\nu}\left(R_{\mu\nu}(\Gamma)+\frac{2}{3}W_{[\mu,\nu]}
\right)\; ,
\end{equation}
where $R_{\mu\nu}(\Gamma)$ is the generalized Ricci tensor, $W_{\mu}$ is
a Lagrange multiplier and $\Gamma$ refers to the torsion-free connection
$\Gamma^{\lambda}_{\mu\nu}$. Square brackets denote anti-symmetrization
and we use units where $G=c=1$ throughout. The vacuum field equations that
follow from (\ref{NGT}) are
\begin{eqnarray}
&& g_{\mu\nu,\sigma} - g_{\rho\nu} {\Gamma}^{\rho}_{\mu\sigma} -
g_{\mu\rho} {\Gamma}^{\rho}_{\sigma\nu} = 0 , \\ \nonumber \\
&& {(\sqrt{-g}g^{[\mu \nu]})}_{ , \nu} = 0 , \\ \nonumber \\
&& R_{\mu \nu}(\Gamma) = \frac{2}{3} W_{[\nu , \mu]} . 
\end{eqnarray} 
The most general, spherically symmetric metric for NGT was found by
Papapetrou\cite{pap} to be
\begin{equation} \label{metric}
g_{\mu\nu}\! = \!\left( \begin{array}{cccc}
\gamma(r)& w(r) & 0 & 0 \\ \nonumber \\
-w(r) & -\alpha(r) & 0 & 0 \\ \nonumber \\
0 & 0 & -r^{2} & f(r)\sin\theta\\ \nonumber \\
0 & 0 & -f(r)\sin\theta& -r^{2}\sin^{2}\theta \end{array} \right)\; .
\end{equation}
In this paper we shall take the NGT charge to be zero which demands $w(r)=0$.
Papapetrou gave a special, unphysical solution to field equations for this
metric which had $f(r)=Ar^2$. This solution is discarded since the resulting
metric is not of Minkowskian form for large $r$. While studying the general
wave behaviour of NGT\cite{cm}, we found that asymptotically Minkowskian
solutions could be found with $f(r)=const.$, since this corresponds to $1/r^2$
fall-off in orthonormal coordinates. The static spherically symmetric solutions
found by Wyman\cite{wy}, Bonnor\cite{bon} and Vanstone\cite{van} for the field
equations of Unified Field Theory share this asymptotic behaviour\cite{pierre}.
This lead us to adapt their solutions to NGT, and we found that the resulting
solution proved to be free of singularities.

The metric functions are given by
\begin{eqnarray}
\gamma&=&e^{\nu}\; , \\ \nonumber \\
\alpha&=&{m^2(\nu ')^2 e^{-\nu} (1+s^2) \over (\cosh(a\nu)-\cos(b\nu))^2}\; ,\\
\nonumber \\
f&=&{2m^2e^{-\nu}(\sinh(a\nu)\sin(b\nu)+s(1-\cosh(a\nu)\cos(b\nu)) \over
(\cosh(a\nu)-\cos(b\nu))^2} \; ,
\end{eqnarray}
where
\begin{equation}
a=\sqrt{{\sqrt{1+s^2}+1 \over 2}}\; , \hspace{0.5in}
b=\sqrt{{\sqrt{1+s^2}-1 \over 2}}\; ,
\end{equation}
prime denotes differentiation with respect to $r$, and $\nu$ is given implicitly
by the relation:
\begin{equation}\label{impl}
e^{\nu}(\cosh(a\nu)-\cos(b\nu))^2{r^2 \over 2m^2}=\cosh(a\nu)\cos(b\nu)
-1+s\sinh(a\nu)\sin(b\nu) \; .
\end{equation}
For small $r$ this relation has multiple branches of solutions for
$\nu$, but only one of these branches matches onto the unique solution
of $\nu$ for large $r$. We shall only be interested in the unique inversion
of (\ref{impl}) which yields an asymptotically flat spacetime.

Unfortunately, we can only invert (\ref{impl}) analytically for
$r/m << 1$ and $m/r << 1$ and we must resort to numerical methods to establish
the intermediate behavior.

Starting in familiar territory, we find for $m/r <<1$ and $s<<1$ that the
metric takes the near-Schwarzschild form:
\begin{eqnarray}
\gamma&=&1-{2m \over r}+{s^2 m^5 \over 15 r^5}+{4 s^2 m^6 \over 15 r^6}
        + \dots \; , \\ \nonumber \\
\alpha&=&1+{2m \over r}+{4m^2 \over r^2}+{8m^3 \over r^3}+{(16-2s^2/9)m^4
        \over r^4} + \dots \; , \\ \nonumber \\
f&=&{sm^2 \over 3}+{2sm^3 \over 3r}+{6sm^4 \over 5r^2} + \dots \; ,
\end{eqnarray}
where the higher order terms in $m/r$ include higher powers of $s$ also.
We see that for large $r$ the NGT corrections to the Schwarzschild solution
are small for any $s<1$, and can be arbitrarily small if $s$ is close to
zero. Clearly, experimental predictions such as the bending of light by
the sun and the perihelion advance of Mercury will only be minutely affected
by having $s$ non-zero.

Near the origin we can develop expansions where $r/m <<1$ and $0<s<<1$. The
leading terms are
\begin{eqnarray}
\gamma&=&\gamma_{0} +{\gamma_{0}(1+{\cal O}(s^2)) \over 2s}
\left({r \over m}\right)^2+{\cal O}((r/m)^4)\; , \label{g}\\ 
\nonumber \\ \nonumber \\
\alpha&=&{4\gamma_{0}+{\cal O}(s^2) \over s^2}\left({r \over m}\right)^2
+{\cal O}((r/m)^4)\; , \label{a} \\ \nonumber \\ \nonumber \\
f&=&m^2\left(4-{s\pi \over 2}+s^2+{\cal O}(s^3)\right)
+{s+s^2\pi/8+{\cal O}(s^3) \over 4}r^2+{\cal O}(r^4)\; ,\label{f}
\end{eqnarray}
where $\gamma_{0}$ is given by
\begin{equation}
\gamma_{0}=\exp{\left(-{\pi \over s}-2+{\cal O}(s)\right)} \; .
\end{equation}
The above equation clearly illustrates the non-analytic nature of the limit
$s \rightarrow 0$ in the strong field region. We shall postulate that $s$ is
always strictly greater than zero, and refer to this new phenomenology for NGT
as Non-Singular Gravity or NSG. 

Our numerical results confirm
that all the expansions we have exhibited produce excellent approximations to
the exact solution in their respective regions of validity.

\begin{figure}[h]
\vspace{60mm}
\includegraphics{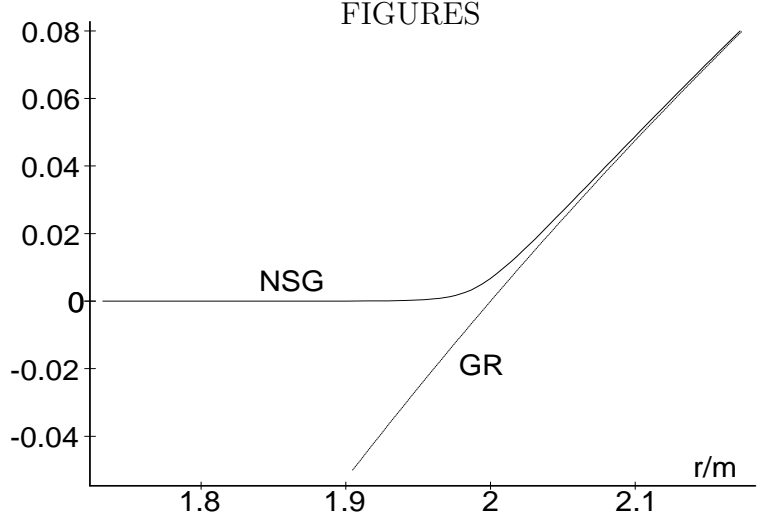}
\vspace{5mm}
\caption{$\gamma(r)$ in the region around $r=2m$ for both GR ($s=0$) and
NSG with the choice $s=0.1$.}
\end{figure} 

Figure 1. shows how $\gamma$ behaves in the region where neither of our
expansions are valid, for the choices $s=0$ (GR) and $s=0.1$. We see that for
NSG the redshift is finite between any two points in the spacetime. The
maximum redshift is between $r=0$ and $r=\infty$, and is given by
\begin{equation}
z=\exp\left({\pi \over 2s}+1+{\cal O}(s)\right)-1 \; .
\end{equation}
Put another way, we find that a timelike Killing vector at spatial infinity
remains timelike throughout the spacetime - our solution is free of event
horizons.

\begin{figure}[h]
\vspace{70mm}
\includegraphics{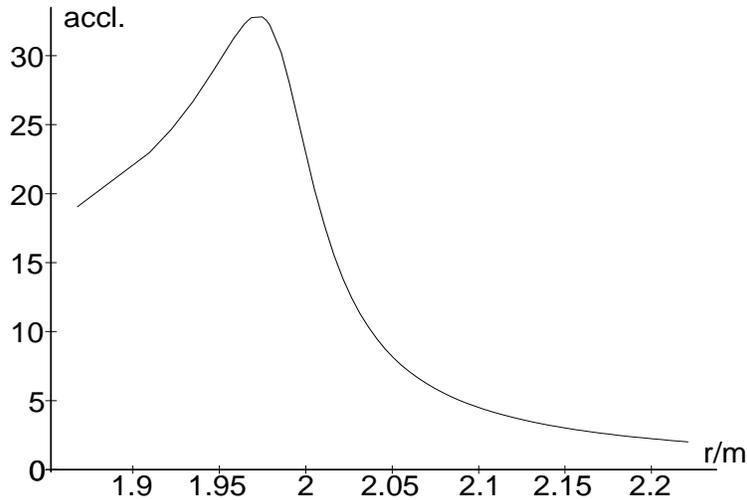}
\vspace{5mm}
\caption{The acceleration of an intially static observer in the region near
$r=2m$ for NSG with $s=0.1$ and $m=1$.}
\end{figure} 

Figure 2. shows the inward radial acceleration of a static observer near
$r=2m$ for the choice $s=0.1$. Unlike GR, we see that the acceleration is
finite at $r=2m$. The radial acceleration in NGT is given by the familiar
expression $a^{r}=1/2(\ln\gamma)'=\nu '/2$. Near $r=0$ we find
\begin{equation}
a^{r}={1+{\cal O}(s^2) \over 2sm^2}r+{\cal O}(r^3/m^4) \; ,
\end{equation}
while for large $r$ we have
\begin{equation}
a^{r}={m \over r^2}+{2m^2 \over r^3}+{4m^3 \over r^4}
+{8m^4 \over r^5}+{(16-s^2/6)m^5 \over r^6} +\dots \; .
\end{equation}
At large distances we have a new, repulsive, $1/r^6$ contribution to the
gravitational ``force'' which is of little interest. In contrast, the
situation near $r=0$ is dramatically different from GR, since the ``force''
vanishes in NSG whereas it is infinite in GR. The shape of the acceleration
curve for initially static observers in NSG is reminiscent of the acceleration
curves one finds for extended matter distributions such as planets or
stars. The acceleration goes roughly as $1/r^2$ outside the body and
tails off smoothly to zero inside the body. We attribute this to the fact
that in NSG the degrees of freedom associated with the skew field $f$
are able to distribute the energy of the point source over an extended region.
Figure 3. shows a plot of the invariant
$F=(g_{[\theta\phi]}g^{[\theta\phi]})^{1/2}=f/\sqrt{f^2+r^4}$. For large $r$
we find $F\sim sm^2/3r^2$, while at the origin $F=1$.

\begin{figure}[h]
\vspace{70mm}
\includegraphics{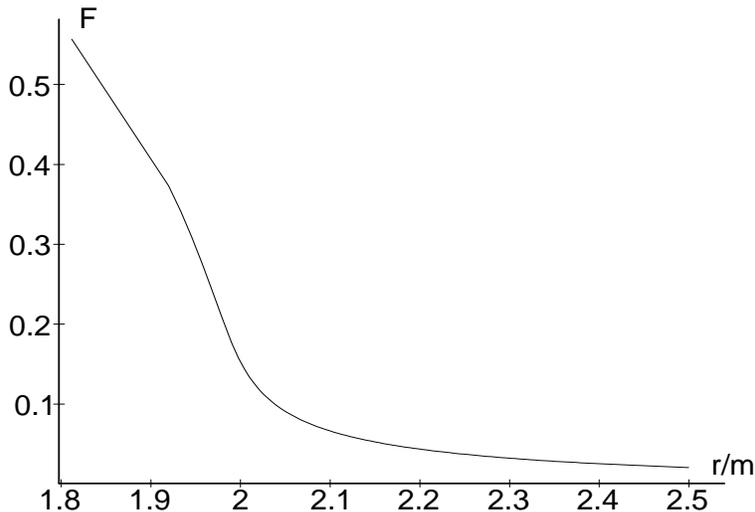}
\vspace{5mm}
\caption{The invariant skew potential near r=2m for NSG with the choice $s=0.1$}
\end{figure} 

We see that $F$ is negligible outside of $r=2m$, while below $r=2m$ it quickly
rises to 1. The radial dependence exhibited by $F$ is similar to
the energy density profile of a compact object with radius $r=2m$, and
suggests the mechanism by which the singularities can be avoided.

The vanishing of the radial acceleration at $r=0$ leads us to expect that the
curvature is non-singular at the origin. This expectation is confirmed by an
explicit calculation of the generalized curvature invariants. NGT generally has
more curvature invariants than GR. However, the fact that $R_{[\mu\nu]}=0$ for
the solution we are studying reduces the number that needs to be calculated to
four. Direct numerical evaluation shows that the curvature is non-singular
for all $r$. Since there is a coordinate singularity in $\alpha$ at $r=0$, it
is also important to establish analytically that there is no curvature
singularityat $r=0$, rather than relying on our numerical results. Using the
expansions (\ref{g}), (\ref{a}) and (\ref{f}), we found all the curvature
invariants to be finite. A representative invariant is the generalized
Kretschmann scalar, which is given to leading order in $s$ by
\begin{eqnarray}
K&=&R^{\mu\nu\kappa\lambda}R_{\mu\nu\kappa\lambda} \;  \\ \nonumber \\
 &=&{\exp(\pi / s+2) \over 2sm^4}\; , \hspace{0.2in} {\rm at}\; r=0\; ,\\
     \nonumber \\
 &=&{48m^2 \over r^6}\; , \hspace{0.2in} r\rightarrow \infty \; .
\end{eqnarray}
All the non-vanishing curvatures are of the form $1/m^2$ at the origin and
$m/r^3$ for large $r$, which is to be expected on dimensional grounds. These
results are again reminiscent of some extended energy distribution.
For example, the non-vanishing curvature invariants at the
center of a star are proportional to the local energy density $\rho$.
Consistent with this picture we see that the curvature is smaller for larger
values of $s$ since for large $s$ the skew field is better able to smooth
out the would-be singularity at $r=0$.

While all the results we have presented suggest that NSG sector of NGT
is indeed a non-singular theory of gravity, there is still much to be understood.
An obvious question is whether the spacetime we have described would
ever be reached as the endpoint of gravitational collapse. Since gravity
tends to turn off at very short distances, it is perhaps more likely that
some kind of super-dense object might form with the metric we have described
as its exterior spacetime. Observationally it would be difficult to
distinguish such objects from black-holes. Looking beyond the limitations
of a static spherically symmetric solution, we need to understand the role
played by the $s$ parameter in more general settings. Is $s$ related to a new
fundamental constant which controls the cross-coupling of the skew and
symmetric parts of the geometry, or is it just a parameter which varies
from one solution to the next? At the linearized level, we know\cite{cm} that
$s$ can be considered as a coupling constant between the three divergence-free
parts of $g_{[\mu\nu]}$ and the rest of the $g_{\mu\nu}$'s, but this
understanding is of little use in the strong field regime where these degrees
of freedom are most important. Another interesting question is whether a
similar smoothing effect by the skew metric can eliminate cosmological
singularities such as the Big Bang. For now we will have to be content with
a non-singular analog of the Schwarzschild geometry.

\section*{Acknowledgements} 
This work was supported by the Natural Sciences and Engineering
Research Council of Canada. One of the authors (NJC) is
grateful for the support provided by a Canadian Commonwealth
Scholarship. We thank P. Savaria,  M. Clayton, B. Holdom and
J. Levin for helpful discussions.

\end{document}